\documentstyle[buckow1,12pt]{article}
\begin{document}

\rightline{HUB-EP-98/08}
\rightline{hep-th/9801154}

\def\titleline{Matrix Theory and the Six Torus
\footnote{Based on
a talk given at the ``31st International Symposium Ahrenshoop
on the Theory of Elementary Particles''
Buckow, September 2-6, 1997.}
  }

\def\authors{
Ilka Brunner , Andreas Karch
}

\def\addresses{
Institut f\"ur Physik, Humboldt-Universit\"at, \\
Invalidenstr. 110, D-10115 Berlin\\
}

\def\abstracttext{
We review the problems associated with Matrix
compactifications on $T^6$.
}

\makefront

\section{Introduction}
Matrix theory claims to be a non-perturbative formulation
of M-theory. However it is formulated in a background
dependend way and hence compactifications pose a serious
problem.

At the heart of the matrix construction lies the fact, that in the
special reference frame chosen several degrees of freedom decouple
\cite{BFSS,DLCQ}.
Basically one considers a compact light-like circle \cite{DLCQ} 
of radius $R$
with $N$ units of momentum $1/R$ around the circle. In the limit
that $N$ and $R$ both go to infinity this goes over to the
so called infinite momentum frame (IMF)
of \cite{BFSS}, which loosely speaking is the reference frame of an observer
boosted infinitely with respect to the experiment along the 11
direction.

Due to the infinite boost, 
in this frame $p_{11}$ of all the constituents 
is much bigger than any other scale in the problem.
The energy of a particle becomes
$$
E=\sqrt{p^2_{11}+p^2_{perp}+m^2} \rightarrow |p_{11}| + \frac{p_{perp}^2+m^2}
{2|p_{11}|}.$$
The Hamiltonian responsible for propagation in this frame
is 
$H=E-p_{11}.$
From this we see, that only modes with positive $p_{11}$ contribute to the
dynamics. In this case the Hamiltonian is governed by
the finite piece
$$H= \frac{p_{perp}^2+m^2}{2|p_{11}|}$$
while all the other modes lead to a Hamiltonian which goes to infinity with
growing $p_{11}$. Therefore they lead to very rapid oscillations and can be neglected.
This decoupling mechanism is still valid when we compactify more
spacelike directions. The main task is to identify these remaining
degrees of freedom.

\section{The matrix description}

To analyse the dynamics of M-theory in the discrete lightcone gauge,
one can use IIA/M-theory duality to map this system to a better understood
IIA setup. One then identifies the degrees of freedom that
survive as carriers of positive $p_{11}$ as D0 branes \cite{BFSS}.
The same philosophy suggest, that using T-duality Matrix theory
on the $T^p$ is described by the dynamics of D$p$ branes.
We will review this kind of construction in the following sections,
especially in view of the still puzzling $T^6$ compactification.
Recently it was shown by Seiberg \cite{seib}
that this identification
can basically be derived from the assumptions
of M-theory/IIA duality and Lorentz invariance of M-theory by
just noting that the compact lightlike circle is Lorentz
equivalent to a particular limit of a compactification
on a vanishing space-like circle.

\subsection{The inifinite boost limit}

In order to identify the surviving degrees of freedom a la \cite{seib}
 one proceeds
in two steps. One notes that the funny frame we are interested
is really equivalent to a vanishing {\it spacelike} 
circle $R$ and one
rescales all length and energy scales involved in order to keep
the relevant Hamiltonian $H$ for the surviving degrees of freedom
finite. These scales are the 11d Planck length $l_p$,
the radii of the transverse compact directions $L_i$ and of
course $R$.
In terms of these quantities we can express $p_{11}$, $m$ and $p_{perp}$
in the expression for the energy of the relativistic excitation
as
\begin{eqnarray*}
m &=& \tilde{m}/l_p \\
p_{11} &=& N/R \\
p_{perp} &=& M_i/L_i 
\end{eqnarray*}
where $\tilde{m}$, $N$ and $M_i$ are fixed dimensionless numbers. The
first relation just expresses the fact that we want to measure
energies in terms of the 11d Planck scale, while the other two
are just quantization conditions for momenta along the compact directions.

Now one wants to take the limit in such a way that for
$R \rightarrow 0$
$$H=E-p_{11} =\frac{p_{perp}^2+m^2}{2|p_{11}|} \sim 
\frac{1/L_i^2+1/l_p^2}{1/R}$$
is constant. Thus the limit of infinite boost parameter amounts to
taking
\begin{equation}
\label{IMF}
R \rightarrow 0, \; \; L_i \rightarrow 0, \; \;
l_p \rightarrow 0  
\end{equation}
$$
l_p^2/R =f = \mbox{fixed},  \; \;  \;   L_i/l_p=D_i =\mbox{fixed} 
$$

\subsection{The type IIA string theory perspective}

After one identified the limit that corresponds to an infinite
boost one can analyze this 
limit from the IIA perspective. By considering the dynamics
of D0 branes (which after all are the carriers
of positive $p_{11}$) in the IIA setup in this particular limit
one therefore derives the matrix description for various string 
compactifications. The parameters of the
corresponding IIA theory
are
$$
\begin{array}{ccc}
g_s^2 = R^3/ l_p^3 &
l_s^2 = l_p^3/R &
L_i = L_i.
\end{array}
$$
We see that the limit (\ref{IMF})
corresponds to $g_s \rightarrow 0$, $M_s=1/l_s \rightarrow \infty$
and $L_i \rightarrow 0$.
If we have no compact transverse dimensions we recover the original
DLCQ version of the matrix description. The full quantum theory
of the sector with $N$ units of momentum $p_{11}$ is described by the
dynamics of $N$ D0 branes in the limit that the string coupling is zero,
the Planck and the string scale go to infinity while the gauge coupling
of the quantum mechanics on the D0 wordvolume is fixed.

Since (\ref{IMF}) involves taking $L_i$ to zero the
original IIA picture is not a good description anymore once
we compactify transverse dimensions.
We should perform a T-duality transformation along
the transverse compact directions, thereby mapping the D0
partons into D$d$ branes, where $d$ denotes the number of compact
transverse dimensions. Since the generalization of T-duality
to arbitrary manifolds is problematic, we will restrict ourselves
in the rest of this work to the case of tori. This way
we will derive the matrix description of the DLCQ of M-theory on $T^d$.
It is given by D$d$ branes wrapping
a torus with radii $\Sigma_i$ in IIB (IIA) string theory for $d$ odd (even), in
the following limit for the 10d string coupling $g_s$, string scale
$M_s$ and 10d Planck scale $M_p$. $g_{YM}$ denotes the coupling constant
of the effective SYM on the D$d$ worldvolume. The following formulas can
be simply obtained by applying the usual T-duality.
$$
\begin{array}{cc}
\begin{array}{ccc}
g_s^2 &=& \frac{l_p^{3d-3}}{R^{d-3} V^2} \\
M_s^2 &=& \frac{R}{l_p^3} \\
M_P^8 &=& \frac{V^2 R^{d+1}}{l_p^{3d+9}} \\
\end{array} &
\begin{array}{ccc}
\Sigma_i &=& \frac{l_p^3}{R L_i} \\
g^2_{YM} &=& \frac{l_p^{3d-6}}{R^{d-3} V} 
\end{array}
\end{array}
$$
where $V= \prod_{i=1}^d L_i$.
From this we can read off how the various parameters behave in the
limit (\ref{IMF}).
$$
\begin{array}{cc}
\begin{array}{ccc}
g_s^2 &\sim& \left \{ \begin{array}{ll}
  0 & \mbox{for  }d<3 \\
  \mbox{finite} & \mbox{for  } d=3 \\
  \infty & \mbox{for  } d> 3
\end{array} \right .\\
M_P^8 &\sim& \left \{ \begin{array}{ll} 
  \infty & \mbox{for  }d<7 \\
  \mbox{finite} & \mbox{for  } d=7 \\
  0 & \mbox{for  } d> 7
\end{array} \right .\\
 \\
\end{array} & \begin{array}{ccc}
M_s^2 &\sim& \infty \mbox{  for all  } d \\
\Sigma_i &\sim& \mbox{finite for all  } d \\
g^2_{YM} &\sim& \mbox{finite for all  } d
\end{array}
\end{array}
$$
Note that this time the sides of the torus stay finite, so we actually
found a good string theory description. Also the gauge coupling of the theory
on the brane is kept at a finite value, so we are really left with
an interacting theory to describe the dynamics of the M-theory setup.
For $d \leq 3$ the string coupling is finite or vanishes, while
Planck and string scale go to infinity. So we can
identify the wordvolume theory on the D$d$ by the usual
string theory techniques. This way one obviously reproduced and hence
derived the usual SYM on the dual torus description.

For $d=4,5,6$ the string coupling blows up while the Planck
scale still goes to infinity. One may hope that one can
use strong/weak coupling dualities to find an appropiate description
in a weakly coupled theory. We will review them in the following
section on a case by case basis.

At least starting from the $T^7$ one won't be able to decouple the bulk
gravity since the Planck scale stays finite. 
In the
infinite boost limit several degrees of freedom decouple. The 
method outlined above allowed us to 
determine which dynamical degrees of freedom
remain. It turns out that for small tori indeed all the
bulk modes decouple, so that the matrix description of M-theory
on $T^d$ is described by a $d+1$ dimensional theory. Even though
for $d\geq7$ still many bulk degrees of freedom decouple (for 
example all the massive string modes, since $M_s$ is sent to infinity)
at least bulk gravity remains! The matrix description of
M-theory in the DLCQ on those higher tori is given by a 10 dimensional
theory! The same turned out to be the case for $d=6$ \cite{seibsethi}.

\section{The four-, five- and sixtorus}

{\bf The fourtorus:}

According to our derivation above the DLCQ of M-theory on the
$T^4$ is given by a IIA D4 brane where we take
the string coupling to infinity, the string scale to infinity
and the 10d Planck scale to infinity, keeping $g^2_{YM}=l_s g_s$, the
gauge coupling of the 4+1 SYM on the D4 fixed. The D4 at infinite
coupling is better thought of as the M-theory 5 brane. Translating
the above limit in M-theory language we find that we are interested in
the worldvolume theory of the M5 in the limit that we take the
radius of the 11th dimension and also the 11d Planck scale to infinity.
This worldvolume theory is known to be given by the (2,0) fixed point
in 6 dimensions. One thus derived the Berkooz-Rozali-Seiberg \cite{t4}
description
of M-theory on the $T^4$.

\noindent
{\bf The fivetorus:}

For the fivetorus we find the theory of D5 branes of IIB at infinite
string coupling, sending $M_s$ and $M_p$ also to infinity,
where as shown above the gauge coupling of the D5 worldvolume
theory stays fixed.
IIB string theory offers us the possibility to map this
to the theory of NS5 branes at zero string coupling. The above limit maps
to $M_p \rightarrow \infty$, $g_s \rightarrow 0$, keeping $M_s$ and
hence the gauge coupling of the NS5 wordvolume gauge theory fixed.
This is precisely Seiberg's realization of the 6 dimensional
little string theory, which he used as the matrix description of
M-theory on $T^5$ \cite{t5}. 

At this point a comment about the BPS solutions of the theory is in order.
Seiberg identified the 16 BPS states preserving 1/2 of the original
32 supercharges, transforming under the $SO(5,5,Z)$ U-duality
group of M-theory on $T^5$. They are bound states of the NS5 brane
with D1, D3 or D5 branes. The energy of these states is given
by  
$$
E=\sqrt{(T_{NS5} V_{NS5})^2 + (T_{D} V_{D})^2} 
$$
where $T_{brane}$ and $V_{brane}$ denote the volume and the tension of the
brane respectively. This is the T-dual version of the formula
$
E=\sqrt{p_{11}^2 + m^2}
$
for the case without transverse momentum. The $p_{11}$ momentum
modes got mapped to $T_{ND5} V_{NS5}$, since this is what the D0 branes,
the carriers of positive $p_{11}$, are mapped to under the ST$^5$-duality
we used. $m^2$ is the mass of the wrapped D branes.
Boosting to the IMF now amounts to taking the
$g_s \rightarrow 0$ limit and switching to a light cone Hamiltonian
$H=E-p_{11}.$
In our case this amounts to taking
$$H=\lim_{g_s \rightarrow 0} \left ( 
\sqrt{(T_{NS5} V_{NS5})^2 + (T_{D} V_{D})^2}
-T_{NS5} V_{NS5} \right )
$$
for the light cone energy of these objects, which is
the relation Seiberg used to get the excitations of
the little string theory. We see again
that this procedure is just equivalent to identifying
the degrees of freedom that survive the infinite boost limit.

Since all the time we are working at finite $R$ we should really see
the U-duality group of M-theory on $T^6$, which is the discrete version
of $E_6$. Under the $SO(5,5,Z)$ subgroup the 27 of $E_6$ decomposes
as
$27 \rightarrow 10 + 16 +1. $
In addition to the 16 states identified above, there are the 10 states
corresponding to longitudinal membranes and fivebranes
(that is branes wrapping $R$), which are momentum and winding modes in 
the little string theory. They correspond to bound states at threshold,
so in this case
$
E=p_{11} + m
$
and hence
their lightcone energy $H=E-p_{11}=m$ is equal to their mass in the IIB string
theory. These states preserve only one quarter of the supersymmetry.
This is due to the fact that in the particular reference frame
we chose only half
of the supercharges are linearly realized. While the transverse branes
break the half that's not visible anyway, they still appear as
1/2 SUSY bound states, while the longitudinal states break another half and
we are left with only 1/4 linearly realized supercharges.

The 27th state is the fundamental carrier of $p_{11}$ itself, the wrapped NS5
brane. Its space-time mass by construction is $1/R$.
Indeed the full U-duality multiplet for the $T^6$ can be found in 
accordance with the fact that $R$ is finite.
\newline
\noindent
{\bf The sixtorus:}

Applying the above procedure to M-theory on the $T^6$, one can
similarly derive that the DLCQ matrix model of this theory
is given by the worldvolume theory of IIA D6 branes at infinite
coupling (sending $M_p$, $M_s$ to infinity, holding the
gauge coupling on the D6 and hence the eleven dimensional Planck
scale fixed). This is the limit proposed in \cite{kk6}.
There it was also found that this description
yields the right moduli space and BPS states.

It was however shown by Seiberg and Sethi \cite{seibsethi} that the
worldvolume theory of the D6 does not decouple from the bulk
fields in this limit. Since we sent the 10d Planck scale $M_p$ to infinity
we decoupled all the 10d gravitons, but the gravitons associated
with the 11th dimension of M-theory that opens up in the 
infinite coupling limit (the D0 branes from the IIA point of view)
become massless in this limit. The coupling of these
excitations to the D6 worldvolume is governed by the 11d Planck scale,
which is kept fixed in the limit that's forced upon us by 
(\ref{IMF}). We thus find that similar as in the case for $T^{\geq 7}$
the matrix description of the DLCQ is not a d+1 dimensional
theory but involves some of the bulk modes of the full type II
theory.

This conlusion can also be reached by looking again at the
BPS states we should see. We are still working at finite $R$
and thus should find the discrete $E_7$ of $T^7$ compactifications.
The 56 of $E_7$ decomposes under the $E_6$ as
$ 56 \rightarrow 27 + \bar{27} + 1 +1 $
In \cite{kk6} the 27 longitudinal and 27 transverse branes were again
identified as bound states of the $p_{11}$ carriers (in this case the D6)
and other branes. One of the singlets is again the
wrapped D6 brane itself and correponds to the space-time state
with energy $1/R$. The 56th state has space-time mass $\frac{R^2 V}{l_p^9}$
and corresponds to the KK6 associated with the compact $R$. But
this state is precisely mapped to a D0 brane in the IIA setup describing
the $T^6$ compactification. These thus can not decouple from
the matrix theory, since they are the required missing BPS state.

By now there seem to be two lines of thought about how to live with
this puzzle. On one hand one might just appeal to the magic of large $N$.
The DLCQ description of M theory on higher tori has all the problems
mentioned. But once we go back to the original proposal of 
\cite{BFSS}, that is go to infinite $N$, all these problems
might go away. The D0 and the D6 repell each other.
For infinite $N$ this might be strong enough to decouple the bulk.
Since we also got rid of the hidden compact dimension, we no longer
would expect to see the full U-duality of the $T^7$ and thus
the D0 branes no longer have to be there. This option was
for example suggestd in \cite{banksrev}.

The other option would be that what we've seen here is an indication
that M-theory doesn't want to be compactified on $T^{\geq 6}$. This
interpretation recently got some support by the observation
that by replacing the $T^6$ by a Calabi-Yau \cite{kachru}, 
the states that correspond
to those that in our case come from the D0
do not lead to any interaction with gravity.


\renewcommand{\baselinestretch}{1} \normalsize


\begin{thebibliography}{99}

\bibitem{BFSS}
T. Banks, W. Fischler, S.H. Shenker, L. Susskind,
hep-th/9610043.
\bibitem{DLCQ}
L. Susskind,
hep-th/9704080.
\bibitem{seib}
N. Seiberg, 
{\it Phys.Rev.Lett.} {\bf79} (1997), 3577, hep-th/9710009;
A. Sen, 
hep-th/9709220.
\bibitem{seibsethi}
N. Seiberg and S. Sethi,
hep-th/9708085.
\bibitem{t4}
M. Berkooz, M. Rozali, N. Seiberg,  hep-th/9704089;
M. Rozali,
hep-th/9702136.
\bibitem{t5}
N. Seiberg, 
hep-th/9705221.
\bibitem{kk6}
I. Brunner, A. Karch, 
hep-th/9707259; A. Hanany, G. Lifschytz,
hep-th/9708037.
\bibitem{banksrev}
T. Banks, `Matrix Theory', hep-th/9710231
\bibitem{kachru}
S. Kachru, A. Lawrence, E. Silverstein, 
hep-th/9712223.
\end{thebibliography}
\end{document}